\documentclass[cmp1]{svjour1}
\usepackage{enumerate}
\usepackage{dsfont}
\usepackage{amssymb,amsfonts,amsmath}
\newcommand{\dd}{\mathrm{d}}
\newcommand{\ee}{\mathrm{e}}

\newcommand{\rh}{r_\mathrm{h}}
\newcommand{\R}{\mathds R}
\newcommand{\Ie}{I_\varepsilon}
\newcommand{\eps}{\varepsilon}
\newcommand{\norm}[1]{\|#1\|}
\newcommand{\artanh}{\mathrm{artanh}}

\newcommand{\arsinh}{\mathrm{arsinh}}%CARLA
\newcommand{\arcosh}{\mathrm{arcosh}}%CARLA
\newcommand{\sign}{\mathrm{sign}}%CARLA
\begin{document}
\title{A universal inequality for axisymmetric and stationary black holes with
surrounding matter in the Einstein-Maxwell theory}
\titlerunning{A universal inequality for black holes with surrounding
matter}
\author{J\"org Hennig \and Carla Cederbaum \and Marcus Ansorg}
\authorrunning{J. Hennig, C. Cederbaum, M. Ansorg}                 
\institute{Max Planck Institute for Gravitational Physics,
Am M\"uhlenberg 1, D-14476 Golm, Germany
\email{pjh@aei.mpg.de, Carla.Cederbaum@aei.mpg.de, mans@aei.mpg.de}}
\date{}
% Add name of the expert who has communicated your paper
%\communicated{name}
%
\maketitle
\begin{abstract}
We prove that in Einstein-Maxwell theory the inequality
$(8\pi J)^2+(4\pi Q^2)^2 < A^2$ holds for any sub-extremal axisymmetric
and stationary black 
hole with arbitrary surrounding matter. Here $J, Q$, and $A$ are angular
momentum, electric charge, and horizon area of the black hole, respectively.
%\\[2ex] Manuscript date: \today
\end{abstract}
%
%%%%%%%%%%%%%%%%%%%%%%%%%%%%%%%%%%%%%%%%%%%%%%%%%%%%%%%%%%%%%%%%%%%%%%%%%%%%%
\section{Introduction}\label{intro}

For a single rotating, electrically charged, axisymmetric and
stationary black hole in vacuum (described 
by the Kerr-Newman family of solutions), the angular momentum $J$, the electric
charge $Q$, and the horizon area $A$ are restricted by the inequality
\begin{equation}\label{ineq}
  p_J^2+p_Q^2 \le 1 \quad\textrm{with}\quad
  p_J := \frac{8\pi J}{A},\quad
  p_Q := \frac{4\pi Q^2}{A}.
\end{equation}
Equality in \eqref{ineq} holds if and only if the Kerr-Newman black hole
is extremal. That is to say, 
\begin{equation}\label{ineq_strict}
  p_J^2+p_Q^2 < 1
\end{equation}
holds for any non-extremal Kerr-Newman black hole. 

As was shown in \cite{Ansorg}, the equality $p_J^2+p_Q^2=1$
holds more generally in Einstein-Maxwell theory for axisymmetric and
stationary \emph{degenerate}\footnote{Degeneracy of an  axisymmetric and
stationary black hole is defined by vanishing surface gravity $\kappa$.}
black holes \emph{with surrounding matter}. Moreover, it was conjectured
in \cite{Ansorg} that inequality (\ref{ineq}) is still valid if the black hole
is surrounded by matter 
(i.e. if it is not a member of the Kerr-Newman family).

Inequality (\ref{ineq_strict}) was proved in
\cite{Hennig} for axisymmetric and stationary black holes with
surrounding matter in pure Einsteinian gravity (without Maxwell field).
In that article, emphasis was put on ``physically
relevant'' configurations by assuming the black hole to be
\emph{sub-extremal}. This condition requires the existence of trapped
surfaces (i.e. surfaces with a negative expansion of outgoing null
geodesics) in every sufficiently small interior vicinity of the event
horizon, see \cite{Booth}. Here, we consider again sub-extremal
axisymmetric and stationary black 
holes with arbitrary surrounding matter, but provide a proof of
(\ref{ineq_strict}) which is valid in the full Einstein-Maxwell theory. 

The idea of the proof relies on showing that a black
hole \emph{cannot} be sub-extremal for $p_J^2+p_Q^2 \ge 1$. In order to
prove this, 
we study the Einstein-Maxwell equations in a
vicinity of the black hole horizon. It turns out that a reformulation
can be found  
which states that an appropriate functional $I$ (to be
defined below) must always be greater than or equal to $1$. 
In this way, we encounter a \emph{variational problem}, and the
corresponding solution provides a proof  
of inequality (\ref{ineq_strict}). As will be shown below,
this variational problem  
can be treated with methods from the calculus of variations.

This paper is organized as follows. In Sec.~\ref{coords}, we introduce
appropriate coordinates which are adapted to the subsequent analysis.
Moreover, we
list the Einstein-Maxwell equations and the corresponding boundary and
regularity conditions in these coordinates. In Sec.~\ref{calc}, 
we express the ingredients $p_J$ and $p_Q$, which appear in
the inequality (\ref{ineq_strict}), in terms of metric and electromagnetic
potentials. We formulate the variational  
problem mentioned above in Sec.~\ref{reform} and solve it in
Sec.~\ref{solution}.  
Finally, we conclude this paper with a discussion on physical
implications of inequality (\ref{ineq_strict}), see Sec.~\ref{discussion}.
In an appendix, we establish a connection to degenerate black holes.

%%%%%%%%%%%%%%%%%%%%%%%%%%%%%%%%%%%%%%%%%%%%%%%%%%%%%%%%%%%%%%%%%%%%%%%%%%%%%
\section{Coordinate systems and Einstein equations}\label{coords}

Following Bardeen~\cite{Bardeen}, we describe an exterior electrovacuum
vicinity of the black hole\footnote{For a stationary spacetime, the
immediate vicinity of a black hole event horizon must be electrovacuum, see
\cite{Carter} and \cite{Bardeen}.}
in spherical coordinates ($R, \theta, \varphi, t$)
in terms of a Boyer-Lindquist-type\footnote{In the special case
\emph{without} any exterior
matter, i.e. for the Kerr-Newman black hole, we obtain Boyer-Lindquist
coordinates ($r,\theta,\varphi, t$) through a linear transformation
$r=2R+M$, where $M$ is the black hole mass.} line element
\begin{equation}\label{LE}
 \dd s^2 = \hat\mu\left(\frac{\dd R^2}{R^2-\rh^2}+\dd\theta^2\right)
           +\hat u\sin^2\!\theta\,(\dd\varphi-\omega\dd t)^2
           -\frac{4}{\hat u}(R^2-\rh^2)\dd t^2,
\end{equation}
where the metric potentials $\hat\mu$, $\hat u$ and $\omega$ are
functions of $R$ and $\theta$ alone and, where in addition $\hat\mu$ and $\hat
u$ are positive functions. The event horizon $\mathcal H$ is located at
$R=\rh$, $\rh=\textrm{constant}>0$.

The electromagnetic field gives rise to an energy momentum tensor
\begin{equation}
 T_{ij} = \frac{1}{4\pi}\left(F_{ki}F^k_{\ j}
          -\frac{1}{4}g_{ij}F_{kl}F^{kl}\right),
\end{equation}
where, using Lorenz gauge, the electromagnetic field tensor $F_{ij}$
can be written in terms of a potential $(A_i) = (0, 0, A_{\varphi}, A_t)$,
\begin{equation}
 F_{ij}=A_{i,j}-A_{j,i}.
\end{equation}
Note that, like the metric quantities, $A_\varphi$ and $A_t$ also depend on $R$
and $\theta$ only.

In the Boyer-Lindquist-type coordinates, the Einstein-Maxwell equations in electrovacuum are
given by\footnote{Throughout this paper we consider vanishing
cosmological constant, $\Lambda=0$. (Note that inequality
\eqref{ineq} can be violated for $\Lambda\neq 0$.
An example is the Kerr-(A)dS family of black holes, see \cite{Booth}.)}
\begin{align}
 & (R^2-\rh^2)\tilde u_{,RR}+2R\tilde u_{,R}+\tilde u_{,\theta\theta}
   +\cot\theta\,\tilde u_{,\theta}\nonumber\\
 & \quad= 1-\frac{\hat u^2}{8}\sin^2\!\theta
   \left(\omega^2_{,R}+\frac{\omega^2_{,\theta}}{R^2-\rh^2}\right)
   -\frac{1}{\hat u\sin^2\!\theta}\left[(R^2-\rh^2)A^2_{\varphi,R}
   +A^2_{\varphi,\theta}\right]\nonumber\\
 & \qquad -\frac{\hat u}{4}\left[(\Phi_{,R}-A_{\varphi}\omega_{,R})^2
   +\frac{(\Phi_{,\theta}-A_{\varphi}\omega_{,\theta})^2}{R^2-\rh^2}\right],
   \label{E1}\displaybreak[0]\\
%%%%%%%%%%%%%%%%%
 & (R^2-\rh^2)\tilde\mu_{,RR}+R\tilde\mu_{,R}+\tilde\mu_{,\theta\theta}
   \nonumber\\
 & \quad = \frac{\hat u^2}{16}\sin^2\!\theta
   \left(\omega^2_{,R}+\frac{\omega^2_{,\theta}}{R^2-\rh^2}\right)
   -(R^2-\rh^2)\tilde u^2_{,R}
   +R\tilde u_{,R}-\tilde u_{,\theta}(\tilde u_{,\theta}+\cot\theta),
   \label{E2}\displaybreak[0]\\
%%%%%%%%%%%%%%%%%
 & (R^2-\rh^2)(\omega_{,RR}+4\omega_{,R}\tilde u_{,R})
   +\omega_{,\theta\theta}+\omega_{,\theta}(3\cot\theta+4\tilde
   u_{,\theta})\nonumber\\
 & \quad=\frac{4}{\hat u\sin^2\!\theta}\left[(R^2-\rh^2)A_{\varphi,R}
   (\Phi_{,R}-A_{\varphi}\omega_{,R})+A_{\varphi,\theta}
   (\Phi_{,\theta}-A_\varphi\omega_{,\theta})\right],\displaybreak[0]\\
%%%%%%%%%%%%%%%%%
 & (R^2-\rh^2)\left[\Phi_{,RR}-A_\varphi \omega_{,RR}
   +2\tilde u_{,R}(\Phi_{,R}-A_\varphi \omega_{,R})
   -A_{\varphi,R}\omega_{,R}\right]\nonumber\\
 & \qquad + \Phi_{,\theta\theta}-A_\varphi\omega_{,\theta\theta}
   +(2\tilde
   u_{,\theta}+\cot\theta)(\Phi_{,\theta}-A_\varphi\omega_{,\theta})
   -A_{\varphi,\theta}\omega_{,\theta} = 0,\displaybreak[0]\\
%%%%%%%%%%%%%%%%%
 & (R^2-\rh^2)\left[A_{\varphi,RR}-2\tilde u_{,R}A_{,\varphi,R}\right]
   +2RA_{\varphi,R}+A_{\varphi,\theta\theta}
   -(2\tilde u_{,\theta}+\cot\theta)A_{\varphi,\theta}\nonumber\\
 & \quad = \frac{\hat u^2}{4}\sin^2\!\theta
   \left[(\Phi_{,R}-A_\varphi\omega_{,R})\omega_{,R}
   +\frac{\Phi_{,\theta}-A_\varphi\omega_{,\theta}}{R^2-\rh^2}
   \omega_{,\theta}\right].
\end{align}
Here, we have used the dimensionless quantities
\[	
	\tilde u:=\frac{1}{2}\ln\frac{\hat u}{\hat u_\textrm{N}}\,,\quad 
	\tilde\mu:=\frac{1}{2}\ln\frac{\hat\mu}{\hat u_\textrm{N}}\,,
\]
where $\hat u_\textrm{N}$ is the the north pole value of $\hat u$, 
\[
	\hat u_\textrm{N}:=\hat u(R=\rh,\theta=0)\,.
\]
Moreover, we have replaced $A_t$ by the comoving electric potential
\begin{equation}\nonumber
 \Phi=A_t+\omega A_\varphi.
\end{equation}

At the horizon, the metric potentials
obey the boundary conditions (cf. \cite{Bardeen})
\begin{equation}\label{BC}
 \mathcal H:\quad
 \omega=\textrm{constant}=\omega_\mathrm{h},\quad
 \frac{2\rh}{\sqrt{\hat\mu\hat u}} = \textrm{constant}=\kappa,\quad
 \Phi=\textrm{constant}=\Phi_\mathrm{h},
\end{equation}
where $\omega_\textrm{h}$, $\kappa$, and $\Phi_\textrm{h}$
denote the angular velocity of the horizon, the
surface gravity, and the value of the comoving electric potential
at the horizon, respectively.

On the horizon's north and south pole ($R=\rh$
and $\sin\theta=0$), 
the following regularity conditions hold\footnote{Note that \eqref{RC1}
holds on the entire rotation axis.}:
\begin{align}\label{RC}
 & \hat\mu(R=\rh,\theta=0)=\hat u(R=\rh,\theta=0)\nonumber\\
 & \quad = \hat\mu(R=\rh,\theta=\pi)=\hat u(R=\rh,\theta=\pi) 
    = \frac{2\rh}{\kappa},
\end{align}
\begin{equation}\label{RC1}
 A_\varphi(R,\theta=0)=A_\varphi(R,\theta=\pi)=0.
\end{equation}

In the forthcoming calculations we need relations between metric 
and electromagnetic quantities at the black hole horizon $\mathcal H$. 
These are provided by an investigation of the equations \eqref{E1} and
\eqref{E2}. For the evaluation of these equations in the limit $R\to\rh$, we introduce the regular horizon
potentials (cf. \cite{Ansorg})
\begin{equation}
 \hat\omega:=\frac{\omega-\omega_\textrm{h}}{R-\rh},\quad
 \hat\Phi  :=\frac{\Phi-\Phi_\textrm{h}}{R-\rh}.
\end{equation}
from which it follows that
\begin{equation*}
 \lim\limits_{R\to\rh}\frac{\omega^2_{,\theta}}{R^2-\rh^2}
 =\lim\limits_{R\to\rh}\left(\frac{R-\rh}{R+\rh}\hat\omega^2_{,\theta}\right)
 =0,
\end{equation*}
\begin{equation*}
 \lim\limits_{R\to\rh}\frac{(\Phi_{,\theta}-A_\varphi\omega_{,\theta})^2}
 {R^2-\rh^2}
 =\lim\limits_{R\to\rh}\left[\frac{R-\rh}{R+\rh}
  (\hat\Phi^2_{,\theta}-A_\varphi\hat\omega_{,\theta})^2\right]
 =0.
\end{equation*}
Using these relations, we obtain for \eqref{E1} and \eqref{E2} in
the limit $R\to\rh$
\begin{equation}
   2\rh\tilde u_{,R}+\tilde u_{,\theta\theta}
   +\cot\theta\,\tilde u_{,\theta}
  = 1-\frac{\hat u^2}{8}\sin^2\!\theta\,
   \omega^2_{,R}
   -\frac{A^2_{\varphi,R}}{\hat u\sin^2\!\theta}
    -\frac{\hat u}{4}(\Phi_{,R}-A_{\varphi}\omega_{,R})^2,
   \label{E1a}
\end{equation}
\begin{equation}
 \rh\tilde\mu_{,R}+\tilde\mu_{,\theta\theta}
  = \frac{\hat u^2}{16}\sin^2\!\theta\,
   \omega^2_{,R}
   +\rh\tilde u_{,R}-\tilde u_{,\theta}(\tilde u_{,\theta}+\cot\theta).
   \label{E2a}
\end{equation}

%%%%%%%%%%%%%%%%%%%%%%%%%%%%%%%%%%%%%%%%%%%%%%%%%%%%%%%%%%%%%%%%%%%%%%%%%%%%%
\section{Calculation of $p_J$ and $p_Q$}\label{calc}

In order to find suitable
expressions for $p_J$ and $p_Q$, we introduce the following
functions which are defined as follows in terms of the metric 
and electromagnetic quantities at the black hole horizon $\mathcal H$: 
\begin{eqnarray}\label{D1}
 U(x) & := & \frac{1}{2}\ln\frac{\hat u}{\hat u_\textrm{N}}
       \Big|_\mathcal H,\quad
 V(x):=\frac{1}{4}\hat u\,\omega_{,R}\big|_\mathcal H,\\
 \label{D2}
 S(x) & := & \frac{\hat u}{2\sqrt{\hat u_\textrm{N}}}
       (\Phi_{,R}-A_\varphi\omega_{,R})\big|_\mathcal H,\quad
 T(x):=\frac{A_\varphi}{\sqrt{\hat u_\textrm{N}}}\Big|_\mathcal H,
\end{eqnarray}
where $x:=\cos\theta$.

In terms of these quantities we obtain for angular momentum $J$, charge
$Q$ and horizon area $A$ (cf. \cite{Ansorg}):
\footnote{Note that $m^i$ denotes the Killing vector with respect to axisymmetry.}
\begin{align}
 J =\, & \frac{1}{8\pi}\oint_\mathcal H(m^{i;j}+2m^k A_k F^{ij})
       \dd S_{ij}\nonumber\\  
   =\, & -\frac{1}{4}\int\limits_0^\pi\hat u\left[\frac{\hat u}{4}\omega_{,R}
       \sin^2\!\theta-A_\varphi(\Phi_{,R}-A_\varphi\omega_{,R})\right]
       \Big|_\mathcal H\sin\theta\,\dd\theta\nonumber\\
   =\, &  -\frac{\hat u_\textrm{N}}{4}
        \int\limits_{-1}^1[V\ee^{2U}(1-x^2)-2ST]\dd x,\displaybreak[0]\\
 Q =\, & -\frac{1}{4\pi}\oint_\mathcal H F^{ij}\dd S_{ij} = 
       -\frac{1}{4}\int\limits_0^\pi
       \hat u(\Phi_{,R}-A_\varphi\omega_{,R})\big|_\mathcal H
       \sin\theta\,\dd\theta\\
   =\, & -\frac{\sqrt{\hat u_\textrm{N}}}{2}\int\limits_{-1}^1 S\,\dd x
   \displaybreak[0]\\
 A =\, & 2\pi\int\limits_0^\pi\sqrt{\hat\mu\hat u}\big|_\mathcal H
         \sin\theta\,\dd\theta = 4\pi\hat u_\textrm{N}\,.
\end{align}
Here, we have used conditions \eqref{BC} and \eqref{RC}. 
Finally, we arrive at
\begin{align}\label{pJ}
 p_J \equiv\, & \frac{8\pi J}{A} = -\frac{1}{2}\int\limits_{-1}^1
     [V\ee^{2U}(1-x^2)-2ST]\dd x,\\
 p_Q \equiv\, & \frac{4\pi Q^2}{A} = \frac{1}{4}\left(\,\int\limits_{-1}^1
     S\,\dd x\right)^2.\label{pQ}
\end{align}
%%%%%%%%%%%%%%%%%%%%%%%%%%%%%%%%%%%%%%%%%%%%%%%%%%%%%%%%%%%%%%%%%%%%%%%%%%%%%
\section{Reformulation in terms of a variational problem}\label{reform}

As a first step towards the proof of the inequality (\ref{ineq_strict})
for sub-extremal black holes,  
we consider the following lemma.
\begin{lemma}[Characterization of sub-extremal black holes]\label{Lem1}
A necessary condition for the existence of trapped surfaces in the
interior vicinity of the event horizon of an
axisymmetric and stationary charged black hole is
 \begin{equation}\label{lemma}
  \int\limits_0^\pi(\hat\mu\hat
  u)_{,R}\big|_\mathcal{H}\sin\theta\,\dd\theta>0.
 \end{equation}
\end{lemma}
This lemma was originally derived in the setting of
pure Einsteinian gravity (without
Maxwell-field), see \cite{Hennig}. As the corresponding proof presented
in \cite{Hennig} carries over to the full Einstein-Maxwell theory, we may
use the lemma in the forthcoming investigation.

The proof of (\ref{ineq_strict}) relies on showing that for
$p_J^2+p_Q^2 \ge 1$ inequality \eqref{lemma} is violated, which implies
by virtue of lemma \ref{Lem1} a violation of the sub-extremality
condition: 
\begin{equation}\label{implication_1}
	p_J^2+p_Q^2 \ge 1\qquad\Rightarrow\qquad   \int\limits_0^\pi(\hat\mu\hat
	u)_{,R}\big|_\mathcal{H}\sin\theta\,\dd\theta\leq 0.
\end{equation}

Using Einstein equations \eqref{E1a} and \eqref{E2a} together with the boundary
conditions \eqref{BC}, we may rewrite the integrand in
\eqref{lemma} as
\begin{align}
 (\hat\mu\hat u)_{,R}\big|_\mathcal{H} =\, &
 \frac{2(\hat u_\textrm{N})^2}{\rh}\left[1-
 \left(\frac{\hat u\omega_{,R}}{4}\right)^2\sin^2\!\theta
 -\tilde u_{,\theta}(\tilde u_{,\theta}+2\cot\theta)\right.\nonumber\\
 & \quad\left.-\frac{A^2_{\varphi,\theta}}{\hat u\sin^2\!\theta}
 -\frac{\hat u}{4}(\Phi_{,R}-A_\varphi\omega_{,R})^2\right].
\end{align}
Hence we can express \eqref{lemma} in terms of $S$, $T$, $U$,
and $V$:
\begin{equation}\label{lemma1}
 \frac{1}{2}\int\limits_{-1}^{1}\left[(V^2+U'^{\,2})(1-x^2)
 -2xU'+(S^2+T'^{\,2})\ee^{-2U}\right]\dd x<1
\end{equation}
where $':=\dd/\dd x$. With the expressions for $p_J$ and $p_Q$
[see \eqref{pJ}, \eqref{pQ}] we can thus write the implication in
(\ref{implication_1}) as follows: 
\begin{align}\label{imp}
 &\left(\,\int\limits_{-1}^1\left[V\ee^{2U}(1-x^2)-2ST\right]\dd x\right)^2
 +\frac{1}{4}\left(\,\int\limits_{-1}^1 S\dd x\right)^4\ge 4\nonumber\\
 &\Rightarrow\quad \frac{1}{2}\int\limits_{-1}^{1}\left[(V^2+U'{\,^2})(1-x^2)
 -2xU'+(S^2+T'{\,^2})\ee^{-2U}\right]\dd x \ge 1.
\end{align}
In the following we show that this implication holds for all sufficiently regular functions\footnote{A precise statement about the required regularity properties follows below.}
\[
	S, T, U, V:[-1,1]\to\R
\]
which satisfy the boundary conditions
\begin{equation}\label{bound_cond_U_T}
 U(\pm 1) = T(\pm 1)=0.
\end{equation}
The conditions in (\ref{bound_cond_U_T}) follow from \eqref{D1},
\eqref{D2}, \eqref{RC}, \eqref{RC1}.

%Thereby, the Sobolev space $W^{1,2}(-1,1)$ is an appropriate set for the
%definition of the functions since it should contain all $S$, $T$, $U$,
%$V$, which belong to physically relevant black holes, and since the
%left hand sides of both inequalities in \eqref{imp} are well-defined on
%$W^{1,2}(-1,1)$.
In the next step we formulate a variational problem which is a
sufficient criterion for the validity of the implication in
\eqref{imp}. 
Applying the Cauchy-Schwarz inequality to the first inequality in
\eqref{imp}, we obtain
\begin{equation*}
 \left(\sqrt{\int\limits_{-1}^1 V^2(1-x^2)\dd x}
 \sqrt{\int\limits_{-1}^1\ee^{4U}(1-x^2)\dd x}
 +2\left|\,\int\limits_{-1}^1 S\,T\dd x\right|
 \right)^2+\frac{1}{4}
 \left(\,\int\limits_{-1}^1 S\dd x\right)^4\ge 4.
\end{equation*}
With the abbreviations
\begin{equation}
 c_1:=\sqrt{\int\limits_{-1}^1\ee^{4U}(1-x^2)\dd x},\quad
 c_2:=\int\limits_{-1}^1 ST\dd x,\quad
 c_3:=\frac{1}{\sqrt{2}}\int\limits_{-1}^1 S\dd x,
\end{equation}
this inequality leads to the estimate
\begin{equation}\label{estimate}
 \int\limits_{-1}^1V^2(1-x^2)\dd x \ge M_2^2,
\end{equation}
where
\begin{equation}
 M_1:=\max\left\{0,4-c_3^4\right\},\quad
 M_2:=\max\left\{0,\frac{\sqrt{M_1}-2|c_2|}{c_1}\right\}.
\end{equation}
Using \eqref{estimate} in order to replace the term
$\int V^2(1-x^2)\,\dd x$
in the second inequality in \eqref{imp}, it follows immediately that
\begin{equation}\label{functional}
 I[S, T, U]:=\frac{1}{2}\int\limits_{-1}^1[U'{\,^2}(1-x^2)-2xU'(x)
             +(S^2+T'{\,^2})\ee^{-2U}]\dd x+\frac{M_2^2}{2}\ge 1
\end{equation}
is a sufficient condition for the validity of the implication
in \eqref{implication_1}. We summarize this result in the following lemma.
\begin{lemma}[Variational problem]\label{Lem2}
 The inequality $p^2_J+p_Q^2<1$ holds for
 any sub-extremal axisymmetric and stationary charged black hole
 with surrounding matter provided that the inequality
 \begin{equation}\label{lemma2}
  I[S, T, U] \ge 1 
 \end{equation}
is satisfied for all $S\in L^2(-1,1)$, $T, U\in W_0^{1,2}(-1,1)$.
\end{lemma}

\emph{Remark:} The Lebesgue and Sobolev spaces $L^2$ and $W^{1,2}_0$ contain
all functions $S$ and $T$, $U$, respectively, that arise in the
physical situation above.

With this lemma, we have reduced inequality \eqref{ineq_strict} to
the \emph{variational problem} of calculating the minimum of $I[S,T,U]$ and
showing that this is greater than or equal to $1$. In the next section,
we solve this problem with methods from the calculus of variations.

%%%%%%%%%%%%%%%%%%%%%%%%%%%%%%%%%%%%%%%%%%%%%%%%%%%%%%%%%%%%%%%%%%%%%%%%%

\section{Solution of the variational problem}\label{solution}

\subsection{An approximating family of functionals}\mbox{}\\[1.5ex]
Analyzing the functional $I$ proves difficult as the factor $1-x^2$
is singular at the boundary $x=\pm1$, cf. definition of $I$ in
\eqref{functional}. We therefore approximate it by
a family of slightly modified functionals
$I_{\varepsilon}$ which are conducive to analysis using techniques of the
calculus of variations. We work on the Hilbert space
\begin{equation}\label{X}
X:=(L^2\times W^{1,2}_0\times W^{1,2}_0)(-1,1)
\end{equation}
endowed with the
inner product
$$\left\langle(S,T,U),(\tilde{S},\tilde{T},\tilde{U})\right\rangle
 :=\int\limits_{-1}^1\left[S\tilde{S}+T'\tilde{T}'
 +U'\tilde{U}' (1+\varepsilon-x^2)\right]\dd x$$
depending on a fixed $\varepsilon>0$. Recall that this inner product
is equivalent to the ordinary one by the fundamental theorem of
calculus. Moreover, we have 

       \begin{proposition}[Theorem 2.2 in Buttazzo-Giaquinta-Hildebrandt
         \cite{Buttazzo}]\label{prop}
          On any bounded interval $J\subseteq\R$, $W^{1,2}(J)\hookrightarrow
          C^0(\overline{J})$ compactly. Moreover, the fundamental theorem of
          calculus holds in $W^{1,2}(J)$.
       \end{proposition}

For $\eps\ge0$,       
we consider the functional $I_\varepsilon:\,X\to\R$ given by
\begin{equation}\label{functional1}
 I_\varepsilon[S,T,U]:=\frac{1}{2}\!\int\limits_{-1}^{1}
   \!\left[U'^{\,2}(1+\varepsilon-x^2)-2xU'+(S^2+T'^{\,2})\ee^{-2U}\right]
   \!\dd x +\frac{M_2^\varepsilon[S,T,U]^2}{2}
\end{equation}
where the auxiliary functionals
$M_2^\varepsilon, M_1, c_1^\varepsilon, c_2, c_3:\,X\to\R$
are defined by
\begin{equation}\label{auxI}
 c_1^\varepsilon[S,T,U]:=\sqrt{\int\limits_{-1}^1\ee^{4U}(1+\varepsilon-x^2)
 \dd x}, \quad
 c_2[S,T,U]:=\int\limits_{-1}^1 ST\, \dd x,
\end{equation}
\begin{equation}\label{auxII} 
 c_3[S,T,U]:= \frac{1}{\sqrt{2}}\int\limits_{-1}^1 S\, \dd x, \quad
 M_1[S,T,U]:=\max\left\{0,4-(c_3[S,T,U])^4\right\},
\end{equation}
and
\begin{equation}\label{auxIII}
 M_2^\varepsilon[S,T,U]
 :=\max\left\{0,\left(\frac{\sqrt{M_1}-2|c_2|}{c^\eps_1}\right)[S,T,U]\right\},
\end{equation}
respectively. All of these functionals can easily be seen to be
well-defined, and all auxiliary functionals are weakly continuous on $X$
by Poincar\'e's inequality. Also, $c^\eps_1$ is positive and both $M_1$,
$M^\eps_2$ are non-negative.

We now show that for $\eps>0$ there exists a global minimizer
$(S,T,U)\in X$ for $\Ie$ and study its value $\Ie[S,T,U]$.
Following this investigation, we take the
limit $\varepsilon\to0$ and see that the claim of lemma \ref{Lem2}
follows. 

\vspace{-1.5ex}
\subsection{Existence and characterization of the minimizer}
\mbox{}\\[1.5ex]
Now let $\eps>0$ be fixed.
$\Ie$ then has the following properties:
\begin{enumerate}[(i)]
\item
$\Ie$ is \emph{bounded from below}. Using $0\le
\left(\frac{x}{\sqrt{1+\eps-x^2}}-U'(x)\sqrt{1+\eps-x^2}\right)^2=
\frac{x^2}{1+\varepsilon-x^2}-2xU'(x)+U'^{\,2}(x)(1+\varepsilon-x^2)$ we
conclude that
$$I_\varepsilon[S,T,U]\ge -\frac{1}{2}\int_{-1}^{1}
\frac{x^2}{1+\varepsilon-x^2}\,\dd x =:C(\eps)> -\infty$$
for any $(S,T,U)\in X$. 

\item
 $\Ie$ is \emph{coercive with respect to the weak topology on
 $X$}. Indeed, applying the Cauchy-Schwarz inequality to $\int_{-1}^{1}
 xU'(x)\dd x$, we obtain that
 $$\Ie[S,T,U] \ge \frac{1}{2}\norm{U}^2-C(\eps)\norm{U}$$
 for any $(S,T,U)\in X$ with
 $C(\eps)>0$. Hence, for every $P\in\mathds R$ there exists
 $Q_P\in\mathds R$ such that $\Ie[S,T,U] \ge P$ whenever
 $\norm{(S,T,U)}\ge Q_P$. This is equivalent to coercivity of the
 functional $\Ie$ with respect to the weak topology on $X$, where both
 the norm $\norm{\cdot}$ and the weak topology refer to the inner product
 defined above. 
 
\item
 The functional $\Ie$ is {\it sequentially lower semi-continuous (lsc) with respect to the weak topology on $X$}.
To see this, recall that lower semi-continuity is additive and that the
first terms can be dealt with by standard theory (see
e.g. \cite{Yosida}), and use proposition \ref{prop} as well as the
Lipschitz continuity of $\exp$ on bounded intervals. For the last term,
the weak continuity of the auxiliary functionals yields the claim. 
\end{enumerate}

We are now in a position to show \emph{existence of a global minimizer for $\Ie$}:\\
As we have seen in (i), $\Ie$ is bounded from below on $X$. We can hence choose a minimizing sequence $(S_k,T_k,U_k)\in X$ which must be bounded by coercivity (ii) and thus has a weakly converging subsequence by Hilbert space techniques (theorem
of Eberlein-Shmulyan \cite{Yosida}) tending to a limit
$(S^\ast,T^\ast,U^\ast)$. Lower semicontinuity as in (iii) then gives us
$\Ie[S^\ast,T^\ast,U^\ast]=\inf\{\Ie[S,T,U]\,|\, (S,T,U)\in X\}$ and
thus asserts that $(S^\ast,T^\ast,U^\ast)$ is a global minimizer.
However, $\Ie$ is not Fr\'echet-differentiable
at $(S,T,U)\in X$ with $c_2[S,T,U]=0$ and $c_3^4[S,T,U]=4$
due to the
\emph{maximum}-terms in the definitions of $M^\eps_i$ ($i=1,2$).
It is consequently
impossible to derive Euler-Lagrange equations for $\Ie$ directly.

To circumvent this problem, we introduce the constraints
$c_i^\eps=\textrm{constant}$ ($i=1,2,3$) and use the method of Lagrange
multipliers to minimize $\Ie$ under these constraints. This leads to a
Fr\'echet-differentiable functional on every class $\mathcal K$
with fixed values of
$c_i^\eps$. Moreover, the asserted global minimizer $(S^*, T^*, U^*)$
also minimizes $\Ie$ in its class $\mathcal K^*$
which induces conditions on the
constants specifying $\mathcal K^*$ and explicit expressions for
the related Lagrange multipliers.   

\vspace{-1.5ex}
\subsection{The Euler-Lagrange equations}\mbox{}\\[1.5ex]
Setting $c_i^\ast:=c_i^\varepsilon[S^\ast,T^\ast,U^\ast]$ and
$M_j^\ast:=M_j^\varepsilon[S^\ast,T^\ast,U^\ast]$ $(i=1,2,3;\, j=1,2)$,
the class $\mathcal K^*$ containing the global minimizer $(S^*, T^*,
U^*)$  is characterized by 
$$\mathcal{K}^\ast:=\left\{(S,T,U)\in X\, | \,
c_i^\varepsilon[S,T,U]=c_i^\ast \;(i=1,2,3)\right\}.$$
In this class, $\Ie$ can be evaluated as follows
$$\Ie[S,T,U]=\frac{1}{2}\int\limits_{-1}^{1}\left[U'^{\,2}(1+\varepsilon-x^2)-2xU'+(S^2+T'^2)\ee^{-2U}\right]\,
\dd x   +\frac{(M_2^\ast)^2}{2}.$$
%In $\mathcal{K}^\ast$, $\Ie$ thus
%depends exponentially on $U$ and quadratically on $S$, $T'$, and $U'$,
%only.
By the theory of Lagrange multipliers, for each minimizer
$(S,T,U)$ of $\Ie$ in the class $\mathcal{K}^\ast$, there is
$(\lambda_1, \lambda_2, \lambda_3)\in\R^3$ such that $(S,T,U,\lambda_1,
\lambda_2, \lambda_3)\in X\times\R^3$ is a critical point of the
functional $J_{\eps}^\ast:\,X\times\R^3\to\R$ given by 
\begin{eqnarray*}
 J_\varepsilon^\ast[S,T,U,\lambda_1, \lambda_2,
 \lambda_3]&:=&\frac{1}{2}\int\limits_{-1}^{1}
 \left[U'^{\,2}(1+\varepsilon-x^2)-2xU'+(S^2+T'^{\,2})\ee^{-2U}\right]\dd x\\ 
&&+ \lambda_1\left((c_1^\varepsilon[S,T,U])^2-(c_1^\ast)^2\right)
  + \lambda_2\left(c_2[S,T,U]-c_2^\ast\right)\\
&&+ \sqrt{2}\,\lambda_3\left(c_3[S,T,U]-c_3^\ast\right),
\end{eqnarray*}
which is well-defined and indeed sufficiently smooth by proposition
\ref{prop}. In other words, there is
$(\lambda_1^\ast,\lambda_2^\ast,\lambda_3^\ast)\in\R^3$ such that
$(S^\ast,T^\ast,U^\ast)$ satisfies 
\begin{align}\nonumber
0 = &
\int\limits_{-1}^{1}\left[U'\psi'(1+\varepsilon-x^2)-x\psi'\right]\dd x
+\int\limits_{-1}^{1}
\left[S\rho+T'\varphi'-(S^2+T'^{\,2})\psi\right]\ee^{-2U}\,\dd x\\ 
& +4\lambda_1^\ast\int\limits_{-1}^{1}\ee^{4U}\psi
(1+\varepsilon-x^2)\,\dd
x+\lambda_2^\ast\int\limits_{-1}^{1}\left(S\varphi+T\rho\right)\dd
x+\lambda_3^\ast\int\limits_{-1}^{1}\rho\,\dd x\label{ELJ} 
\end{align}
for all $(\rho,\varphi,\psi)\in X$. This can be restated by saying that
$(S^\ast,T^\ast,U^\ast)$ is a weak solution of 
\begin{eqnarray}\label{ELJI}
 0 & = & -U''(1+\varepsilon-x^2)+2xU'+1-(S^2+T'^{\,2})\ee^{-2U}
 +4\lambda_1^\ast(1+\varepsilon-x^2)\,\ee^{4U}\quad\\\label{ELJII}
 0 & = & -T''+2U'T'+\lambda_2^\ast S\ee^{2U}\\\label{ELJIII}
 0 & = & S+(\lambda_2^\ast T+\lambda_3^\ast)\ee^{2U}\\\nonumber
 0 & = & T(\pm1)=U(\pm1)
\end{eqnarray}
on $(-1,1)$. Any weak solution $(S,T,U)\in X$ of the system \eqref{ELJI}, \eqref{ELJII}, and \eqref{ELJIII} can be shown to be
smooth and to satisfy the equations strongly via a bootstrap argument: For all
$(\rho,\varphi,\psi) \in X$, we can rewrite \eqref{ELJ} as
\begin{eqnarray*}
 0 & = & \int\limits_{-1}^{1}
 \left[U'(1+\varepsilon-x^2)-x
 +\int\limits_{-1}^x(S^2+T'^{\,2})\,\ee^{-2U}\dd t\right.\\
 && \left.
  -4\lambda_1^\ast\,\int\limits_{-1}^{x}\ee^{4U}(1+\varepsilon-t^2)\dd t\right]
 \psi'\,\dd x,\\
 0 & = & \int\limits_{-1}^{1}
 \left[T'\ee^{-2U}-\lambda_2^\ast\int\limits_{-1}^x S\,\dd
 t\right]\varphi'\,\dd x,\\
 0 & = & \int\limits_{-1}^{1} \left[S\ee^{-2U}
 +\lambda_2^\ast T+\lambda_3^\ast\right]\rho\,\dd x,
\end{eqnarray*}
where we used integration by parts and proposition \ref{prop}. By the fundamental lemma of the calculus of variations,
there are constants $a, b\in\R$ such that the equations
\begin{eqnarray}\label{ELJa}
 a & = & U'(x)(1+\varepsilon-x^2)-x
 +\int\limits_{-1}^x(S^2+T'^{\,2})\,\ee^{-2U}\dd t\nonumber\\
 && -4\lambda_1^\ast\,\int\limits_{-1}^{x}
 \ee^{4U}(1+\varepsilon-t^2)\,\dd t,\\\label{ELJb}
 b & = & T'(x)\,\ee^{-2U(x)}-\lambda_2^\ast\int\limits_{-1}^x S\,\dd t,
 \\ \label{ELJc}
 0 & = & S(x)\,\ee^{-2U(x)}+\lambda_2^\ast\,T(x)+\lambda_3^\ast
\end{eqnarray}
hold almost everywhere on $(-1,1)$. Solving iteratively for $T'$, $U'$,
and $S$, we deduce the respective smoothness of $S$, $T$, and $U$ up to
the boundary by a bootstrap argument (similar to p.~462 in \cite{Evans})
using propostion \ref{prop} in every step. Differentiating equations
\eqref{ELJa} and \eqref{ELJb}, we get validity of \eqref{ELJI}
and \eqref{ELJII} in the strong sense.
\emph{In particular, $(S^\ast,T^\ast,U^\ast)$ is
a smooth classical solution of the Euler-Lagrange equations of
$J_\varepsilon^\ast$ with $(\lambda_1, \lambda_2,
\lambda_3)=(\lambda_1^\ast, \lambda_2^\ast, \lambda_3^\ast)$.} 

\vspace{-1.5ex}
\subsection{Solution of the Euler-Lagrange equations}\mbox{}\\[1.5ex]
Let us now determine the minimizer $(S,T,U):=(S^\ast,T^\ast,U^\ast)$
explicitly, dropping the asterisk in what follows for ease of
notation. $S$ can obviously be expressed as 
\begin{equation}\label{Sexpl}
S(x)=-[\lambda_3+\lambda_2\,T(x)]\,\ee^{2U(x)}
\end{equation}
by equation \eqref{ELJIII}. Inserting this expression into
\eqref{ELJII}, we get the equation 
\begin{equation}\label{ELT}
0=T''-2U'T'+\lambda_2\,(\lambda_3+\lambda_2\,T)\,\ee^{4U},
\end{equation}
a linear ODE of second order for $T$ for given $U$. To solve
\eqref{ELT}, consider two separate cases: 
\begin{enumerate}[(i)]  
 \item Assume $\lambda_2=0$. Then \eqref{ELT} reduces to $T''=2UT'$ which
 has the general solution $T(x)=a\int_{-1}^x\ee^{2U(t)}\,\dd t+b$ with
 $a, b\in\R$, so that $T(\pm1)=0$ induces $T\equiv0$. 
 \item Assume now $\lambda_2\neq0$. In this case, \eqref{ELT} has the
 general solution 
 \begin{equation}\label{Texpl}
  T(x)=\frac{\lambda_3}{\lambda_2}\,\left[a\sin\left(\lambda_2\,
  \int\limits_{-1}^x\ee^{2U(t)}\,\dd t\right)+b\cos\left(\lambda_2\,
  \int\limits_{-1}^x\ee^{2U(t)}\,\dd t\right)-1\right]
 \end{equation}
 with $a, b\in \R$. For $\lambda_3\neq 0$,
 inserting the boundary values $T(\pm1)=0$ gives us $b=1$ and
 \begin{equation}\label{A}
  a=\frac{1-\cos\left(\lambda_2\,\int\limits_{-1}^1\ee^{2U(t)}\,\dd
  t\right)}{\sin\left(\lambda_2\,\int\limits_{-1}^1\ee^{2U(t)}\,\dd
  t\right)}=\pm\sqrt{\frac{1-\cos\left(\lambda_2\,\int\limits_{-1}^1
  \ee^{2U(t)}\,\dd t\right)}{1+\cos\left(\lambda_2\,
  \int\limits_{-1}^1\ee^{2U(t)}\,\dd t\right)}}. 
 \end{equation}
\end{enumerate}
The task of determining $U$ remains to be completed.
To this end, set
\begin{equation}\label{gamma}
 \gamma:=-\left[S(x)^2+T'(x)^2\right]\ee^{-4U(x)}\leq0
\end{equation}
and observe that $\dd\gamma/\dd x=0$, so that $\gamma$ is a
non-positive constant. Moreover, from the explicit expressions obtained
for $S$ and $T$, we see that $\gamma=-\lambda_3^2\,(1+a^2)$ where, as
defined above, $a=0$ if $\lambda_2=0$ and $a$ is as in
\eqref{A} otherwise.
 
Recall that $(S,T,U)$ is a global minimizer of $\Ie$. Although $\Ie$ is
not globally Fr\'echet-differentiable w.r.t. $S$ and $T$, it can
straightforwardly be shown that it is continuously
Fr\'echet-differentiable w.r.t. $U$. We thus deduce via integration
by parts and by the fundamental lemma of the calculus of variations that
$$0=-U''(1+\varepsilon-x^2)+2xU'+1-(S^2+T'^{\,2})\ee^{-2U}
-\frac{2M_2^2}{c_1^2}\,(1+\varepsilon-x^2)\,\ee^{4U}.$$
Comparing this equation with \eqref{ELJI}, we obtain the explicit
expression
\begin{equation}\label{lambda1}
 \lambda_1=-\frac{M_2^2}{2c_1^2}\leq0.
\end{equation}
Moreover, the Euler-Lagrange equation \eqref{ELJI} for $U$, which can
now be written as
\begin{equation*}
  0 = -U''(1+\varepsilon-x^2)+2xU'+1+\gamma\ee^{2U}
      +4\lambda_1(1+\varepsilon-x^2)\,\ee^{4U},
\end{equation*}
has an integrating factor and leads to the first order ODE 
\begin{eqnarray}\label{F}
F & := & -(1+\varepsilon-x^2)^2U'^{\,2}+2x(1+\varepsilon-x^2)U'
      +2\lambda_1\,\ee^{4U}(1+\varepsilon-x^2)^2\nonumber\\
  &&  -x^2+\gamma\,(1+\varepsilon-x^2)\,\ee^{2U}\equiv\textrm{constant},
\end{eqnarray}
because
\begin{eqnarray*}
 F'(x) & = & 2[(1-x^2)U'(x)-x]\Big[-U''(1+\varepsilon-x^2)+2xU'+1
             +\gamma\ee^{2U}\\
       &   & +4\lambda_1(1+\varepsilon-x^2)\,\ee^{4U}\Big] = 0.  
\end{eqnarray*}

We now proceed to calculate $U$.
Substituting $W(x):=(1+\varepsilon-x^2)\,\ee^{2U(x)}>0$ on $[-1,1]$,
equation \eqref{F} can be reformulated to say 
\begin{equation}\label{W}
 \frac{W'}{2W}=\pm\frac{\sqrt{2\lambda_1 W^2+\gamma W-F}}{1+\varepsilon-x^2},
\end{equation}
which in particular implies $F\leq0$ as both $\lambda_1, \gamma\leq0$
and $W>0$ by definition. We would like to divide by the square root on
the right hand side and integrate the equation.
We must first find out where the
zeros of $W'$ can lie, if they exist at all.
A careful discussion of the ODE \eqref{W} refering to \eqref{ELJI}, the
boundary values $W(\pm 1)=\eps$, and using
the fact that we are discussing the class $\mathcal K^*$ containing the global
minimizer, shows that $W'$ has exactly one zero $\tilde x\in(-1,1)$
and that $W''(\tilde x)<0$. 
Moreover, from this discussion we obtain $F<0$ and the fact that
$\lambda_1$ and $\gamma$ cannot vanish simultaneously. 

Integrating
\eqref{W} on both $\left[-1,\tilde{x}\right)$ and 
$\left(\tilde{x},1\right]$ and using $W''(\tilde{x})<0$ to determine the
correct sign on each interval we obtain
$$ W_{\pm}(x) = \frac{2F}{\gamma-\sqrt{\gamma^2+8\lambda_1 F}
                  \cosh y_{\pm}(x)},\quad
   y_{\pm}(x) := \frac{2\sqrt{-F}}{\sqrt{1+\varepsilon}}\,
                  \artanh(\frac{x}{\sqrt{1+\varepsilon}})\pm C,$$
where
$$C = -\frac{2\sqrt{-F}}{\sqrt{1+\varepsilon}}\,
         \artanh\left(\frac{1}{\sqrt{1+\varepsilon}}\right)
         +\artanh\left(\frac{\sqrt{2\lambda_1\varepsilon^2
         +\gamma\varepsilon-F}}{\sqrt{-F}
         +\frac{\gamma\varepsilon}{2\sqrt{-F}}}\right),$$
$W_-:[-1,\tilde{x})\to\R$, $W_+:(\tilde{x},1]\to\R$.
As the solution $W$ we are looking for is smooth by the above and agrees
with $W_\pm$ where they exist, $W_{-}$ and $W_{+}$ must smoothly fit
together at $\tilde{x}$. Also, the induced functions $U_{-}, U_{+}$ both
smoothly extend to $[-1,1]$ and must agree at 
$\tilde{x}$ to all orders. Moreover, they both solve equation
\eqref{ELJI}.
Thus, Picard's uniqueness theorem (cf.~p.~9
in \cite{Rauch}) tells us they agree on the whole interval $[-1,1]$.
From $W_-(-x)=W_+(x)$ we deduce symmetry of $W$, $\tilde{x}=0$, and $C=0$.

Altogether, we know that $W$ has the following form
\begin{equation}\label{Wexpl}
 W(x)=\frac{2F}{\gamma-\sqrt{\gamma^2+8\lambda_1 F}\cosh y(x)},\quad
 y(x):=\frac{2\sqrt{-F}}{\sqrt{1+\varepsilon}}\,
 \artanh(\frac{x}{\sqrt{1+\varepsilon}}).
\end{equation}
%Moreover, we have
%$C=0$ so that
%$y_1:=y(1)=\artanh(\frac{\sqrt{2\lambda_1\varepsilon^2+\gamma\varepsilon-F}}{\sqrt{-F}+\frac{\gamma\varepsilon}{2\sqrt{-F}}})$.
\vspace{-1.5ex}
\subsection{Estimating the minimal value of $\Ie$}\mbox{}\\[1.5ex]
In order to estimate the value of $\Ie$ at its global minimizer, we use
the fact that 
\eqref{F} allows us to
simplify our expression for $I_\eps$. Using \eqref{lambda1}, we obtain
\begin{equation}\label{Ieps}
 \Ie[S,T,U]=1-\left(\sqrt{1+\varepsilon}
 +\frac{F}{\sqrt{1+\varepsilon}}\right)\,
 \artanh\frac{1}{\sqrt{1+\varepsilon}}. 
\end{equation}
We now intend to estimate $F$ from above
via 
\begin{equation}\label{F1}
 F\leq-(1+\varepsilon)\,\left[1-\frac{2\,\ln\frac{2+\varepsilon}
 {2-\varepsilon}}{\ln\frac{(2+\varepsilon)^2}{\varepsilon}}\right]^2,
\end{equation}
which allows us to conclude that $\liminf\limits_{\eps\to 0}\Ie\ge 1$, see
subsection \ref{limit}.
We prepare this estimate with the study of two auxiliary functions $f$ and
$g$, see below. We then use these functions to obtain
\eqref{F1} in the cases $c_2\neq 0$ and $c_2=0$ (and several subcases),
see subsections \ref{Fall1} and \ref{Fall2}.   

We define
$$ f(\alpha):=\frac{1}{2}M^\eps_2[(1+\alpha)S,T,U]^2,\quad
g(\alpha):=\frac{1}{2}M^\eps_2[S,(1+\alpha)T,U]^2.$$
The function $g:\R\to\R$ can be seen to be
differentiable at $\alpha=0$ and we obtain
\begin{equation}\label{der-g}
g'(0)=-\frac{2\,|c_2|\,M_2}{c_1}.
\end{equation}
As $(S,T,U)$ simultaneously is a minimizer of $\Ie$ and a
critical point of $J_\eps$, it follows from \eqref{ELJ}
on the other hand that
\begin{equation}\label{Oho-g}
0=\int\limits_{-1}^1 T'^{\,2}\,\ee^{-2U}\dd
x+\lambda_2\,c_2=\int\limits_{-1}^1 T'^{\,2}\,\ee^{-2U}\dd x+g'(0).
\end{equation}
We also find that $f:\R\to\R$
is differentiable at $\alpha=0$ unless both $c_2=0$ and
$c_3^4=4$. Recall that this singular case also led us to the introduction of
Lagrange multipliers.
We then have
\begin{equation}\label{der-f}
 f'(0)=\begin{cases}
 g'(0)-\frac{2\,M_2\,c_3^4}{c_1\,\sqrt{M_1}} & \mbox{ if }M_1\neq0\\
   0 & \mbox{ if }M_1=0
 \end{cases},
\end{equation}
unless both $c_2=0$ and $c_3^4=4$.
In addition, it follows as above that 
\begin{equation}\label{Oho-f}
 0=\int\limits_{-1}^1 S^2\ee^{-2U}\dd x+\lambda_2\,c_2+\sqrt{2}\,\lambda_3\,c_3
 =\int\limits_{-1}^1 S^2\ee^{-2U}\dd x+f'(0),
\end{equation}
or equivalently
\begin{eqnarray}\label{Oho-f2}
 0 & = & -\gamma\int\limits_{-1}^1\ee^{2U}\dd x
         -\int\limits_{-1}^1 T'^{\,2}\ee^{-2U}\dd x
         +\lambda_2\,c_2
         +\sqrt{2}\,\lambda_3\,c_3\nonumber\\
   & = & -\gamma\int\limits_{-1}^1\ee^{2U}\dd x
          -\int\limits_{-1}^1 T'^{\,2}\ee^{-2U}\dd x
          +f'(0).
\end{eqnarray}
\vspace{-1.5ex}
\subsection{Estimating the minimal value of $\Ie$: the case $c_2=0$}
\label{Fall1}
\mbox{}\\[1.5ex]
The explicit expression \eqref{der-g} for $g'(0)$ suggests separate
treatment of the cases $c_2=0$ and $c_2\neq 0$. We begin with $c_2=0$.
Four different subcases arise, namely
\begin{enumerate}[(a)]
\item\label{0} $c_3^4=4$
\item\label{a} $c_3=0$
\item\label{b} $c_3\neq0$, $M_1=0$
\item\label{c} $c_3\neq0$, $M_1\neq0$.
%\item\label{d} $c_2\neq0$.
\end{enumerate}
We will find that the last two cases cannot occur in the minimizing
class $\mathcal K^*$. In the first two cases, we will indeed arrive at
estimate \eqref{F1}.  

Let us discuss the singular case \eqref{0} first.
Here, \eqref{Oho-g} implies $T\equiv0$, $c_3\neq0$ assures $S\neq0$ so that we can deduce $\lambda_2=0$ from \eqref{ELJII}. Recall $M_1=M_2=\lambda_1=0$, $\gamma\neq0$. Then \eqref{Sexpl} implies $S=-\lambda_3\,\ee^{2U}$
so that $c_3=-\frac{\lambda_3}{\sqrt{2}}\,\int_{-1}^1\ee^{2U}\,\dd x$. Let us proceed to calculate $\int_{-1}^1\ee^{2U}\,\dd x$.
The boundary condition $W(\pm1)=\varepsilon$ implies
$\gamma=\frac{2F}{\varepsilon(1+\cosh y_1)}$
and we are in a
position to calculate
$$\int\limits_{-1}^1 \ee^{2U(x)}\dd
 x=\int\limits_{-1}^1 \frac{W(x)}{1+\varepsilon-x^2}\dd
 x=-\frac{\sqrt{-F}}{2\gamma}\int\limits_{-y_1}^{y_1}
 \left[1-\tanh^2\!\left(\frac{y}{2}\right)\right]\,\dd
 y=\frac{\varepsilon\sinh(y_1)}{\sqrt{-F}}.$$
Recalling
$\gamma=-\lambda_3^2$, we deduce
$2=c_3^2=2\varepsilon\sinh^2{\left(\frac{y_1}{2}\right)}$
so that $\cosh y_1=\frac{2+\varepsilon}{\varepsilon}$, whence by
definition of $y_1$, $F=-(1+\varepsilon)$ in accordance with
\eqref{F1}.

We now proceed to a discussion of case \eqref{a}. From \eqref{Oho-g} and \eqref{Oho-f}, we get $T\equiv0$ and $S\equiv0$, respectively. This implies $\gamma=0$ so that $\lambda_1\neq0$ by the above. We therefore obtain
$$W(x)=\frac{\sqrt{-F}}{\sqrt{-2\lambda_1}\cosh y(x)}$$
so that the boundary condition $W(\pm1)=\varepsilon$ leads to
$\lambda_1=\frac{F}{2\varepsilon^2\cosh^2 y_1}$. We calculate
$$c_1^2=\int\limits_{-1}^1 \frac{W^2}{1+\varepsilon-x^2}\,\dd
x=-\frac{\sqrt{-F}}{2\lambda_1}\tanh y_1.$$ Recall that in this
particular case also $\lambda_1=-\frac{2}{c_1^4}$ by \eqref{lambda1} so that
$y_1=\arsinh\frac{2}{\varepsilon}$ and we arrive at estimate
\eqref{F1} using
$$\arsinh\, x = \ln(x+\sqrt{x^2+1})\quad\textrm{and}\quad
  \artanh\,x=\frac{1}{2}\ln\frac{1+x}{1-x}.$$

Let us continue with case \eqref{b}. From \eqref{Oho-g} we
get $T\equiv0$, whereas $M_1=0$ implies $M_2=0$ and thus $\lambda_1=0$
by \eqref{lambda1}. On the other
hand, we get $f'(0)=0$ from \eqref{der-f} so that by \eqref{Oho-f2}
we have $\gamma=0$, a contradiction, because we have seen in the
previous subsection that $\lambda_1$ and $\gamma$ cannot vanish
simultaneously. 

Finally, we discuss case \eqref{c}.
As before, we get $T\equiv0$ and thus by \eqref{ELJII}
$\lambda_2=0$ as $c_3\neq 0$ ensures $S\not\equiv 0$. Equation
\eqref{Oho-f} then leads to
$\lambda_3=-\frac{\sqrt{2}c_3^3}{c_1^2}$. From this, we obtain
$\int_{-1}^1\ee^{2U}\,\dd x=\frac{c_1^2}{c_3^2}$ 
where we used \eqref{Sexpl} and $c_3\neq 0$. Also, $\gamma=-\lambda_3^2$
so that 
$\gamma=-\frac{2c_3^6}{c_1^4}$. In particular,
$\Ie[S,T=0,U]=\Ie[0,0,U]+\frac{c_3^4}{2c_1^2}>\Ie[0,0,U]$. This
contradicts $[S,T=0,U]$ being a global minimizer of $\Ie$.
\vspace{-1.5ex}
\subsection{Estimating the minimal value of $\Ie$: the case $c_2\neq 0$}
\label{Fall2}
\mbox{}\\[1.5ex]
Finally let $c_2\neq0$. If $\lambda_1=0$ were
possible, then by \eqref{lambda1} $M_2=0$ so that $g'(0)=0$ and thus
$T\equiv0$ follow from \eqref{der-g}, \eqref{Oho-g}. Equation
\eqref{der-f} then
tells us that $f'(0)=0$ and whence $S\equiv0$, so that also
$\gamma=0$, in contradiction to the above exclusion of
$\lambda_1=\gamma=0$. Thus, $\lambda_1\neq0$ which implies both
$M_2\neq0$ and $M_1\neq0$. 
Using again \eqref{Oho-g}, \eqref{Oho-f}, and \eqref{lambda1}, we obtain: 
\begin{eqnarray}\label{lambda11}
\lambda_1&=&-\frac{(\sqrt{4-c_3^4}-2|c_2|)^2}{2c_1^4}\\\label{lambda2}
\lambda_2&=&-\frac{2(\sqrt{4-c_3^4}-2|c_2|)}{c_1^2}\sign(c_2)\\\label{lambda3}
\lambda_3&=&-\frac{\sqrt{2}(\sqrt{4-c_3^4}-2|c_2|)c_3^3}{\sqrt{4-c_3^4}\,c_1^2}.
\end{eqnarray}
Rewrite $S, T$ in terms of $A(x):=\lambda_2\,\int\limits_0^x
\ee^{2U(t)}\dd t$, $A_1:=A(1)$ and use $W(\pm1)=\varepsilon$, equation
\eqref{gamma}, and our definition of $y_1$ as well as symmetry of $U$
to obtain
\begin{align}
 \label{Tfin}
 T(x) =\, &\frac{\lambda_3}{\lambda_2}
       \left[\frac{\cos A(x)}{\cos A_1}-1\right]\displaybreak[0]\\
 \label{Sfin}
 S(x) =\, &-\lambda_3\frac{\cos A(x)}{\cos
 A_1}\,\ee^{2U(x)}\displaybreak[0]\\
 \label{g}
 \gamma =\, &-\frac{\lambda_3^2}{\cos^2\! A_1}\\
 \nonumber
 c_1^2 =\, & -\frac{\sqrt{-F}}{2\lambda_1}\left[
 \frac{\sqrt{\gamma^2+8\lambda_1 F}\sinh y_1}
 {\sqrt{\gamma^2+8\lambda_1 F}\cosh y_1-\gamma}\right.\\
 &+\left.\frac{2\gamma}{\sqrt{8\lambda_1 F}}\,
 \arctan\left(\frac{\gamma+\sqrt{\gamma^2+8\lambda_1 F}}{\sqrt{8\lambda_1
 F}}\,
 \tanh\frac{y_1}{2}\right)\right]\displaybreak[0]\\
 c_2 =\, & -\frac{\lambda_3^2}{\lambda_2^2}\,
 \frac{A_1-\sin A_1\,\cos A_1}{\cos^2\! A_1}\displaybreak[0]\\
 c_3 =\, & -\sqrt{2}\,\frac{\lambda_3}{\lambda_2}\,\tan A_1\displaybreak[0]\\
 A_1 =\, & \frac{\lambda_2}{\sqrt{-2\lambda_1}}
 \arctan\left(\frac{\gamma+\sqrt{\gamma^2+8\lambda_1 F}}
 {\sqrt{8\lambda_1 F}}\,\tanh\frac{y_1}{2}\right)\displaybreak[0]\\
 \varepsilon =\, & \frac{2F}{\gamma-\sqrt{\gamma^2+8\lambda_1 F}\cosh
 y_1}
 \displaybreak[0]\\
 \label{y1}
 y_1=\, & \frac{2\sqrt{-F}}{\sqrt{1+\varepsilon}}\artanh\frac{1}
 {\sqrt{1+\varepsilon}}.
\end{align}
Now set $\phi:=\arccos\frac{-\gamma}{\sqrt{\gamma^2+8\lambda_1
F}}\in(0,\frac{\pi}{2})$ which is well-defined as $\lambda_1 \cdot F>0$.
Using this new constant, equations \eqref{Tfin} through \eqref{y1} take
on a simpler form. In particular, these equations lead to
$0<|A_1|\leq\phi<\frac{\pi}{2}$ and
\begin{equation*}
 c_1^2 =\frac{4\,\tan\phi\,\cos A_1\,|\sin A_1-A_1 \cos A_1|}
        {\sqrt{-F}\,\sin^4\! A_1},\quad
 c_3=\pm\sqrt{2\,\cos A_1}.
\end{equation*}
For the Lagrange multipliers, we get
\begin{eqnarray*}
\lambda_1&=& \frac{F\,\sin^4\! A_1}{8\tan^2\phi\,\cos^2\! A_1},\\
\lambda_2&=&\frac{4(\sin A_1- A_1\,\cos A_1)}{c_1^2\,\sin^2\! A_1},\\
\lambda_3&=&\mp \frac{4}{c_1^2}\cos^{\frac{3}{2}}\!A_1\,\frac{|\sin A_1 -A_1\,\cos A_1|}{\sin^3\! A_1}.
\end{eqnarray*}

With the above expressions, we obtain
$$\varepsilon=\frac{2\cos A_1\,\sin^2\!\phi}{\sin^2\!
A_1\,\cos\phi\,(\cosh y_1+\cos\phi)}.$$
As $\frac{\sin^2\! x}{\cos x}$ is monotonically increasing on
$\left(0,\frac{\pi}{2}\right)$ and
$|A_1|\leq\phi$, we have $\varepsilon\geq\frac{2}{\cosh y_1+1}$ or in
other words $$y_1\geq\arcosh\left(\frac{2}{\varepsilon}-1\right).$$ This
implies
$$\sqrt{-F}\geq\frac{\sqrt{1+\varepsilon}\,\arcosh\left(\frac{2}{\varepsilon}-1\right)}{2\,\artanh\frac{1}{\sqrt{1+\varepsilon}}},$$
where we have used \eqref{y1}. Recall
$\arcosh\,x=\ln\left(x+\sqrt{x^2-1}\right)$ to deduce \eqref{F1} also
in the discussed case $c_2\neq 0$.
\vspace{-1.5ex}
\subsection{The limit $\eps\to 0$}\label{limit}
\mbox{}\\[1.5ex]
We conclude as promised that for $c_2=0$ cases \eqref{b} and \eqref{c}
cannot apply for the minimizer
$(S,T,U)$, whereas in the remaining cases
\eqref{0} and \eqref{a}, as well as for $c_2\neq 0$, we can
estimate using \eqref{Ieps} and \eqref{F1} that 
\begin{eqnarray}\nonumber
 \Ie[S,T,U]&\geq&
 1+\sqrt{1+\varepsilon}\,\left[\left(1-\frac{2\,
 \ln\frac{2+\varepsilon}{2-\varepsilon}}{\ln\frac{(2+\varepsilon)^2}
 {\varepsilon}}\right)^2-1\right]\,\artanh\frac{1}{\sqrt{1+\varepsilon}}
 \\ \label{Ifin}
 & \geq & 1-2\sqrt{1+\varepsilon}\,\ln\frac{2+\varepsilon}{2-\varepsilon}.
\end{eqnarray}

We now study the limit $\eps\to 0$. For any $(S,T,U)\in\,X$,
$c_2[S,T,U]$, $c_3[S,T,U]$ and thus $M_1[S,T,U]$ are independent of
$\varepsilon$, $\lim_{\varepsilon\to0}c_1^\varepsilon[S,T,U]=c^0_1[S,T,U]$
and thus $\lim_{\varepsilon\to0}M_2^\varepsilon[S,T,U]=M^0_2[S,T,U]$ so
that 
$$
|I[S,T,U]-\Ie[S,T,U]|\leq\frac{\varepsilon}{2}\int\limits_{-1}^1
U'(x)^2\dd x
+\left|\frac{M_2[S,T,U]^2}{2}-\frac{M_2^\varepsilon[S,T,U]^2}{2}\right|\to0$$
as $\varepsilon\to0$, i.e. $\Ie[S,T,U]$ is continuous at $\eps=0$ for
fixed $(S,T,U)$. This finally leads us to an esimate of the original
functional $I$. We obtain
\begin{eqnarray*}
 I[S,T,U] & = &\lim_{\varepsilon\to0}\Ie[S,T,U]\\
 &\geq&\liminf_{\varepsilon\to0}\Ie[S^\ast,T^\ast,U^\ast]\\
 &\geq&\liminf_{\varepsilon\to0}\left(1-2\sqrt{1+\varepsilon}\,
 \ln\frac{2+\varepsilon}{2-\varepsilon}\right)\\ 
 & = & 1,
\end{eqnarray*}
where $(S^*,T^*,U^*)$ denotes the global minimizer of $\Ie$.

This proves the claim of lemma~\ref{Lem2} and therefore the inequality
\eqref{ineq_strict}.
\newline\mbox{}\hfill\mbox{}\qed

Finally, after we have seen that the functional
$I$ has a lower bound of $1$, we can
ask the question of whether there exist functions $S$, $T$,
and $U$ for which $I$ takes on
this value. The investigation of this question
together with a discussion of the meaning of $I$ in the context of
degenerate black holes can be found in appendix~\ref{appendix}. 

%Applying Lemma~\ref{Lem2}, we have shown $p_J^2+p_Q^2<1$ for
%\emph{sub-extremal} black holes.
%%%%%%%%%%%%%%%%%%%%%%%%%%%%%%%%%%%%%%%%%%%%%%%%%%%%%%%%%%%%%%%%%%%%%%%%%
\section{Discussion}\label{discussion}

With techniques from the calculus of variations, we have
shown that the inequality $p_J^2+p_Q^2<1$ holds for
axisymmetric and stationary sub-extremal black holes with surrounding
matter in full Einstein-Maxwell theory.

In particular, we have proved the inequality $I[S,T,U]\ge 1$ for the
functional $I$ defined in \eqref{functional}. As $I$ could not directly be
seen to have a local minimizer, we introduced a family of
approximating functionals $\Ie$ which could be shown to have one.

Together with a theorem for \emph{degenerate} black holes in
\cite{Ansorg}, we can deduce the following.
\begin{theorem}\label{theorem}
 Consider Einstein-Maxwell spacetimes with
 vanishing cosmological constant. Then,
 for every axisymmetric and stationary sub-extremal
 black hole with arbitrary
 surrounding matter we have the inequality
 $$(8\pi J)^2 +(4\pi Q^2)^2 < A^2.$$
 If the axisymmetric and stationary black hole is degenerate,
 the equation
 $$(8\pi J)^2 +(4\pi Q^2)^2 = A^2$$ holds.
\end{theorem}

%Note that here we have used a generalization of the result in \cite{Ansorg}
%which requires weaker assumptions, see appendix~\ref{appendix}.
Observe that the assumptions for the result in \cite{Ansorg} which has
been used here have been weakened, see appendix~\ref{appendix}. 

Theorem \ref{theorem} provides a remarkable relation between the
\emph{geometrical} 
concept of the existence of trapped surfaces and the \emph{physical}
black hole properties described by
rotation rate $p_J$ and charge rate $p_Q$. We see that
``physically reasonable'' (sub-extremal)
black holes cannot rotate ``too fast'' and cannot
be charged ``too strongly''. 

Finally, our results shed new light on the notions of
\emph{sub-extremality} and \emph{extremality} of axisymmetric and
stationary black holes. Any sub-extremal black hole in the sense of
\cite{Booth} (the notion of which we have used throughout this paper)
is also sub-extremal in the sense that $p_J^2+p_Q^2<1$.
In fact, $p_J^2+p_Q^2=1$ holds in the degenerate limit, for which reason
we may call these black holes ``extremal''.

%%%%%%%%%%%%%%%%%%%%%%%%%%%%%%%%%%%%%%%%%%%%%%%%%%%%%%%%%%%%%%%%%%%%%%%%%
\begin{acknowledgement}
We would like to thank Herbert Pfister for many valuable discussions
and John Head for commenting on the manuscript.
This work was supported by the Deutsche
Forschungsgemeinschaft (DFG) through the
Collaborative Research Centre SFB/TR7
``Gravitational wave astronomy'' and by the International Max
Planck Research School for ``Geometric Analysis, Gravitation and String
Theory''. 
\end{acknowledgement}
%%%%%%%%%%%%%%%%%%%%%%%%%%%%%%%%%%%%%%%%%%%%%%%%%%%%%%%%%%%%%%%%%%%%%%%%%
\appendix

\section{Remarks on degenerate black holes}\label{appendix}

In order to discuss extremal black holes (as done in \cite{Ansorg})
and to get an idea of the meaning of the appearing functional $I$,
one can apply similar techniques as used in Sec.~\ref{solution}
to $I$ itself to derive
Euler-Lagrange equations and a complete characterization of the
minimizers of $I$. As minimizers of $I$ need not be limits of minimizers
of $\Ie$, this renewed analysis is necessary.

It turns out that the Euler-Lagrange equations for $S$ and $T$ are just
as before, cf. \eqref{ELJII} and \eqref{ELJIII}. Moreover,
there again exists an integrating
factor for the Euler-Lagrange equation for $U$ leading to
$$
 -1=-(1-x^2)^2U'^{\,2}+2x(1-x^2)U'
      +2\lambda_1\,\ee^{4U}(1-x^2)^2
    -x^2+\gamma\,(1-x^2)\,\ee^{2U},
$$
where $\gamma$ is defined as in \eqref{gamma}.
Introducing $W(x):=(1-x^2)\ee^{2U(x)}$ on the interior $(-1,1)$, we find
that equation \eqref{W} holds on $(-1,1)$ with $\eps=0$ and
$F=-1$. Discussing the radicand in \eqref{W} as before,
we see that it vanishes at at
most one inner point. Assuming non-vanishing of the radicand and
integrating the equation on $(-1+\delta,1-\delta)$ for some $\delta>0$
leads to a contradiction as the unique solution $U$ derived from $W$
diverges as $\delta\to
0$ while we know that the smooth solution $U$ exists on the whole
interval by the same bootstrap argument as sketched above.
Thus we know that there exists exactly one interior zero of
the radicand and we can integrate the equation as before to obtain
\begin{equation}\label{U}
 \ee^{2U(x)} = \frac{2}{(1+\gamma)x^2+1-\gamma},
\end{equation}
and the consistency condition
\begin{equation*}
\gamma^2-8\lambda_1=1, 
\end{equation*}
using the boundary values for $U$, $U(\pm1)=0$.
In other words, $U$ belongs to a
family parametrized by $\gamma\in[-1,0]$.

Proceeding as above, we find that $S$ and $T$ are given by
\begin{eqnarray*}
 S(x) & = &
  \pm\left[(-\gamma)^\frac{3}{2}+\sqrt{1-\gamma^2}|T(x)
 |\right]\ee^{2U}\\
 T(x) & = &
 \pm\sqrt{-\gamma(1-\gamma^2)}\frac{1-x^2}{1-\gamma+(1+\gamma)x^2}
\end{eqnarray*}
with $\gamma\in[-1,0]$. The signs of $S$ and $T$ can be chosen
independently of each other.
It can \emph{a posteriori} be seen that all functions $S$,
$T$, $U$ of this form with $\gamma\in[-1,0]$ in fact satisfy
$$I[S, T,U]=1$$
so that we have identified all minimizers of $I$.
Moreover, one can show that the Lagrange parameters $\lambda_1$,
$\lambda_2$, $\lambda_3$ can explicitly be expressed as
\begin{equation}\label{la123}
 \lambda_1 = -\frac{1}{8}(1-\gamma^2),\quad
 \lambda_2 = \pm\sqrt{1-\gamma^2},\quad
 \lambda_3 = \pm(-\gamma)^\frac{3}{2},
\end{equation}
where again the signs are not correlated. 

By comparison with \cite{Ansorg}, one finds that these are exactly the
functions $S$, $T$, and $U$ arising in the context of
arbitrary degenerate black
holes with surrounding matter\footnote{The parameter
$\alpha\in[-1,1]$ in \cite{Ansorg} is related to $\gamma$ via
$\gamma=-(1-\alpha^2)/(1+\alpha^2)$.}.
Moreover, the differential equations characterizing $S$, $T$, and $U$ in
\cite{Ansorg} are exactly the Euler-Lagrange equations of $I$
derived in this paper where the Lagrange parameters in \eqref{la123}
correspond to the constants appearing in \cite{Ansorg}.

We arrive at two conclusions:
First, our analysis dispenses with additional assumptions made in
\cite{Ansorg}, namely equatorial symmetry and the existence of a
continuous sequence of spacetimes, leading from the Kerr-Newman solution
in electrovacuum to the discussed black hole solution.
The latter was necessary to assure uniqueness (up to a parameter)
of the solution to
the horizon equations in \cite{Ansorg}.

As a matter of fact, any smooth solution of the equations in \cite{Ansorg}
is a minimizer of $I$ as can be seen by solving the equations
as done above and using the relations in \eqref{la123} between $\gamma$
and the Lagrange parameters. 
Thus, any solution of these equations is automatically equatorially
symmetric and of the form assumed in \cite{Ansorg}. Hence, the unnecessary
assumptions of \cite{Ansorg} can be dropped.

Secondly, we see that the functional $I$ plays the role of a
``primitive'' of the Einstein equations on the event horizon of
degenerate black holes: Remarkably,
the Euler-Lagrange equations corresponding to
$I$ lead uniquely to the electromagnetic and metric potentials $S$, $T$,
and $U$ belonging to degenerate black holes.
%%%%%%%%%%%%%%%%%%%%%%%%%%%%%%%%%%%%%%%%%%%%%%%%%%%%%%%%%%%%%%%%%%%%%%%%%%%%%

\end{document}